\documentclass[fleqn,usenatbib]{mnras}

\usepackage{newtxtext,newtxmath}

\usepackage[T1]{fontenc}

\DeclareRobustCommand{\VAN}[3]{#2}
\let\VANthebibliography\thebibliography
\def\thebibliography{\DeclareRobustCommand{\VAN}[3]{##3}\VANthebibliography}

\usepackage{graphicx}	
\usepackage{amsmath}	
\usepackage{soul}
\usepackage{xcolor}
\usepackage{orcidlink} 

\title[Cycle dependence of helioseismic pseudomodes]{Cycle dependence of helioseismic oscillations above the acoustic cut-off frequency}

\author[Kolotkov et al.]
{Dmitrii Kolotkov$^{1, 2}$\thanks{E-mail: D.Kolotkov.1@warwick.ac.uk}\orcidlink{0000-0002-0687-6172},
Anne-Marie Broomhall$^{1}$\orcidlink{0000-0002-5209-9378},
Laura Jade Millson$^{1}$\orcidlink{0009-0003-4254-2676},
Sergey Belov$^{1}$\orcidlink{0000-0002-3505-9542}
\\
$^{1}$Centre for Fusion, Space and Astrophysics, Department of Physics, University of Warwick, Coventry CV4 7AL, UK\\
$^{2}$Engineering Research Institute \lq\lq Ventspils International Radio Astronomy Centre (VIRAC)\rq\rq, Ventspils University of Applied Sciences, Ventspils, LV-3601, Latvia
}

\date{Accepted XXX. Received YYY; in original form ZZZ}

\pubyear{2025}

\begin{document}
\label{firstpage}
\pagerange{\pageref{firstpage}--\pageref{lastpage}}
\maketitle

\begin{abstract}
Helioseismic and recent asteroseismic observations reveal fine structure in the power spectrum with alternating peaks and troughs above the acoustic cut-off frequency. This structure is interpreted as the interference patterns of high-frequency acoustic waves excited in the solar interior and propagating into the atmosphere, known as pseudomodes. Pseudomodes exhibit clear solar-cycle variability, with frequency shifts that occur predominantly in anti-phase with the activity cycle, although the underlying mechanism remains uncertain.
This work investigates how the subsurface excitation source location and the photospheric acoustic cut-off frequency influence the formation, frequency distribution, and solar-cycle variability of pseudomodes.
We employ an analytical Klein-Gordon subsurface cavity model, which is shown to act as an effective Fabry-P\'erot interferometer for high-frequency waves that experience constructive and destructive interference between the source location and the lower turning point. We derive an effective dispersion relation isolating the effects of the source location and photospheric cut-off on the pseudomode frequency.
The model reproduces the observed peak-trough pseudomode spectrum for reasonable parameter values constrained by Bayesian MCMC best-fitting to GONG observations. We also find that solar-cycle-associated 11-year modulations of the source location result in anti-phase pseudomode frequency shifts, whereas similar cyclic variations in the cut-off frequency produce harmonic-dependent behaviour, yielding both in-phase and anti-phase shifts.
As the acoustic cut-off and mode excitation relate to stratification and flows in the solar interior, the results highlight pseudomodes as a powerful diagnostic tool for changes in subsurface solar and stellar structure through the solar cycle.

\end{abstract}




\section{Introduction}\label{sec:intro}
Helioseismology uses waves that propagate through the solar interior to infer conditions beneath the visible surface. The primary focus of helioseismology is trapped acoustic oscillations, known as p modes, which form standing waves and have relatively stable properties and easily observable amplitudes \citep[e.g.][]{2016LRSP...13....2B}. The p modes become standing waves as they are trapped in cavities inside the Sun, where the oscillations are refracted at what is known as the lower turning point and reflected near the surface \citep[e.g.][]{RevModPhys.74.1073}. However, only oscillations with frequencies below the ``acoustic cut-off frequency'' are reflected, thus confining them to the solar interior. Oscillations with frequencies above the acoustic cut-off frequency are free to propagate out into the solar atmosphere. When examining helioseismic observations in frequency space, as is commonly done to highlight the p modes, one might therefore expect the power spectrum to be featureless above the acoustic cut-off frequency. However, clear evidence of structure in this high-frequency region of the spectrum is routinely observed \citep{jefferies1988helioseismology, 1988ApJ...334..510L, 1990LNP...367...87K, 1991ApJ...373..308D, 2005ApJ...623.1215J}, with the peaks commonly referred to as ``pseudomodes'' \citep{1989ApJ...342..558K}. These pseudomodes are believed to occur because of constructive interference between waves that have initially been excited to travel towards the solar core and have undergone refraction at their lower turning point and those waves that initially propagate out towards the surface \citep{1991ApJ...375L..35K, 1998MNRAS.298..464V, 1998ApJ...504L..51G}. This results in an oscillatory-like feature in the power spectrum, where the peaks represent frequencies at which constructive interference occurs and the troughs those frequencies at which destructive interference dominates.

Both p modes and pseudomodes are thought to be excited by turbulent convection in the solar interior \citep[e.g.][]{1977ApJ...212..243G, 1989ApJ...342..558K, 1992MNRAS.255..639B, 1995MNRAS.272..850R, 2005MNRAS.360..859C, 2006A&A...460..183B, 2019ApJ...880...13Z, 2022A&A...664A.164P}. Models of solar acoustic oscillations have inferred that the location of excitation depends on the oscillation frequency (with high frequency modes being excited closer to the surface) and the extent and polarity of the source \citep[e.g.][]{1996ApJ...472..882A, 1999MNRAS.309..761C, 1999ApJ...519..396K, 2000MNRAS.314...75C, 2000ApJ...545L..65K}. However, all of these studies agree that the acoustic source will be near the surface, somewhere between around 100--1000\,km beneath the photosphere \citep[e.g.][]{1994ApJ...428..827K, 1998MNRAS.298..464V, 1999MNRAS.309..761C}, and that the source will be localised in radial extent (e.g. \citealt{1994ApJ...428..827K} suggest that while the radial extent of the source is difficult to constrain, they set an upper limit of $550\,\rm km$). For p modes, such a localised, near-surface excitation source is observationally evident primarily in the form of asymmetric peaks in frequency-power spectra \citep[][and references therein]{1993ApJ...410..829D, 1993A&A...274..935G, 1995MNRAS.272..850R, 1996ApJ...472..882A, 2020A&A...635A..81P}. Such asymmetries are small and, therefore, difficult to measure. For pseudomodes, however, the locations of the peaks and troughs in frequency-power spectra are highly sensitive to the source location. Pseudomodes could, therefore, be a hugely useful tool in constraining the excitation of both solar and stellar acoustic oscillations.

While it is well known that p-mode frequencies shift in-phase with solar activity (\citealt{woodard1985change}; \citealt{palle1989solar}; \citealt{libbrecht1990solar}; \citealt{elsworth1990variation}), solar pseudomode frequencies have instead been shown to vary in anti-phase, decreasing as magnetic activity increases. Early indications of this behaviour from \citet{ronan1994solar} found a negative shift in pseudomode frequencies between low (1988) and high (1991) solar activity, and from \citet{rhodes2011temporal}, who showed that frequency shifts above 5000~$\upmu$Hz alternated from being correlated with magnetic activity to anti-correlated depending on frequency. More recently, \citet{kosak2022multi} analysed variations in pseudo-mode frequencies between 100-day segments over nearly 20 years of data, and found frequencies to shift in anti-phase with the solar cycle for harmonic degrees $4\leq\ell\leq200$. Using spatially resolved data, the anti-correlation between solar pseudomode frequencies and the magnetic activity cycle has also been shown to persist across all latitudes throughout the solar cycle \citep{millson2024latitudinal}. Understanding the nature of the solar-cycle pseudomode variations has become increasingly important recently with the detection of stellar pseudomodes \citep{2015A&A...583A..74J, 2025MNRAS.536.3007M}. Moreover, \citet{2025MNRAS.537.1268M} found that while the frequencies of pseudomodes seem to vary with time, for some stars that variation is in anti-phase with the acoustic p-mode frequency variations, like the Sun. For other stars, however, the pseudomode and p-mode frequencies vary in phase.

\citet{1998MNRAS.298..464V} demonstrated that such an anti-phase pseudomode solar cycle variation can be modelled using a simple acoustic cavity, where the reflectivity at the photosphere is varied through the cycle.
Similarly, \citet{2024MNRAS.533.3387K} used the Klein-Gordon equation to describe acoustic waves, analogous to p modes, trapped in a potential well, where the finite upper boundary of the well was determined by the acoustic cut-off. It was demonstrated that having a finite value for the upper boundary impacts the locations of frequencies of the standing mode solutions to the Klein-Gordon equation, in comparison to a perfectly reflecting photosphere (represented by an infinite upper boundary). Furthermore, the impact of an acoustic cut off was modelled to vary in time by the 11-year modulation of a magnetic field. This was motivated by the work of \citet{2011ApJ...743...99J}, which demonstrated that the observed acoustic cut-off frequency varies in phase with the solar cycle. \citet{2024MNRAS.533.3387K} showed that such an in-phase solar cycle modulation of the acoustic cut off produced an in-phase solar cycle variation in p-mode frequency, matching observations.

In this paper, we use a similar theoretical model to determine what may cause the observed solar cycle variation of pseudomodes. In particular, we examine how the subsurface source location and the photospheric acoustic cut-off frequency influence the pseudomode peaks and troughs in helioseismic power spectra above the acoustic cut-off frequency, and investigate whether variations in these parameters could produce the observed solar-cycle-associated pseudomode frequency shifts.
The paper is structured as follows. Section~\ref{sec:model} describes the Klein-Gordon model of acoustic waves near the cut-off frequency and presents the interferometric properties of high-frequency pseudomodes. Section~\ref{sec:alpha_r0_effects} presents an effective dispersion relation for the pseudomode frequency influenced by the subsurface source location and the photospheric cut-off frequency. In Section~\ref{sec:observ}, we compare the outcomes of our model with observations via best-fitting the observed pseudomode spectrum and reproducing the observed solar-cycle modulation of pseudomode frequencies. The discussion of the obtained results and conclusions are summarised in Section~\ref{sec:disc}.

\section{Subsurface acoustic cavity as a Fabry-P\'erot interferometer}
\label{sec:model}

As in \citet{2024MNRAS.533.3387K}, to describe the dynamics of acoustic waves near the cut-off region, we employ the Klein-Gordon equation,
\begin{equation}\label{eq:KG}
    \frac{\partial ^2\psi}{\partial t^2}
    -\frac{\partial ^2\psi}{\partial r^2} +\omega_\mathrm{ac}^{2}(r)\psi = f(t,r),
\end{equation}
where $\psi(t,r)$ is a particular spherical harmonic of an acoustic perturbation, and $\omega_\mathrm{ac}(r)$ is the radial profile of the acoustic cut-off angular frequency. The term on the right-hand side, $f(t,r)$ describes the subphotospheric wave source. The derivation of Eq.~(\ref{eq:KG}) from a more general hydrodynamic model with gravitational stratification and its applicability for acoustic waves in the solar interior were demonstrated in detail in e.g. \citet{1995MNRAS.272..850R} and \citet{2008SoPh..251..523T}. In particular, \citet{1995MNRAS.272..850R} showed that the propagation of acoustic waves in stellar interiors can be considered nearly vertical and described in 1D by Eq.~(\ref{eq:KG}) for $\ell(\ell+1)/\omega^2 \ll r^2/c_\mathrm{s}^2$ with $\ell$, $\omega$, and $c_\mathrm{s}$ being the wave's spherical degree, frequency, and propagation speed; and $r$ being the radial distance. Traditionally, the above condition is deemed to be fulfilled for spherical degrees $\ell \lesssim 100$. However, if one accounts for the expression of the wave's lower turning point $r_\mathrm{l}^2/c_\mathrm{s}^2(r_\mathrm{l})=\ell(\ell+1)/\omega^2$, it can be reformulated as $r\gg r_\mathrm{l}$. The latter merely means the path of a spherical wave (of any $\ell$) appears nearly vertical if one looks at a distance sufficiently far from the lower turning point.

\begin{figure}
    \centering
    \includegraphics[width=\columnwidth]{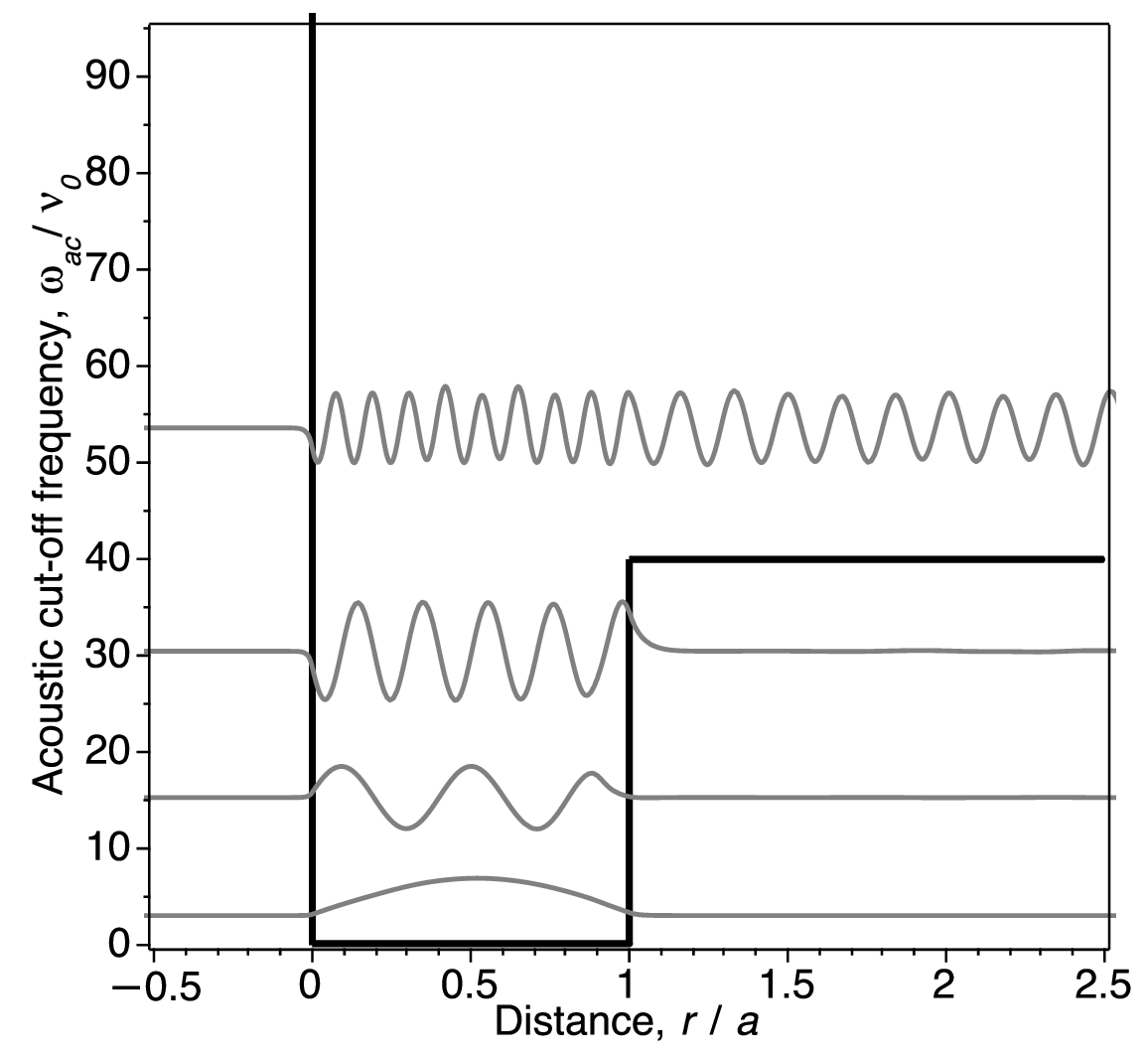}
    \caption{Acoustic cut-off frequency radial profile (the thick black line) given by Eq.~(\ref{eq:cutoff_profile}), normalised to $\nu_0=833$\,$\mu$Hz (corresponding to the 20-min acoustic travel time through the subsurface acoustic cavity of $0.1R_\odot$ depth). The radial distance $r$ is normalised to $a=0.1R_\odot$ corresponding to the acoustic cavity depth of $\ell=100$ modes {with frequencies near the 3-mHz spectral peak}. The height $r/a=1$ is the photospheric level, and $r=0$ mimicks the wave's lower turning point.
    The oscillatory signals (the thin grey lines) illustrate the spatial structure of low-frequency (below the acoustic cut-off, $\alpha=40$ {which is about 5300\,$\mu$Hz if one accounts for normalisation to $\nu_0=833\,\mu$Hz, $\alpha\nu_0/2\pi$)} resonant acoustic waves of radial harmonics $n=1$, 5, and 10 (with $\omega/\nu_0 = 3.06$, 15.3, and 30.48, respectively) trapped within the cavity; and the high-frequency (above the acoustic cut-off) propagating pseudomode with $\omega/\nu_0 = 53.65$. The wave forms are obtained as numerical solutions of the governing Eq.~(\ref{eq:KG}).}
    \label{fig:cutoff_profile}
\end{figure}

In this work, we consider the radial profile of the acoustic cut-off frequency $\omega_\mathrm{ac}(r)$ approximated by a piecewise function (see Fig.~\ref{fig:cutoff_profile}),
\begin{equation}\label{eq:cutoff_profile}
    \omega_\mathrm{ac}(r) = \begin{cases}
			 \alpha, & \text{for $r \ge a$ }\\
             0, & \text{for $0<r<a$} \\
            \gg \alpha, & \text{for $r\le 0$}
		 \end{cases}
\end{equation}
where $a$ is the subphotopsheric acoustic cavity depth, $\alpha$ is the value of the acoustic cut-off angular frequency at the photospheric height $r=a$ \citep[for the Sun,  $\alpha/2\pi$ is around 5\,mHz, see e.g.][]{jimenez2006estimation}, and $r=0$ corresponds to the lower turning point, where the refraction of acoustic waves is effectively mimicked by the reflection from an infinite potential barrier. We use $a=0.1R_\odot$ (the acoustic cavity depth for $\ell=100$ modes {with frequencies near the 3-mHz spectral peak) for all radial harmonics (i.e. the model does not account for a frequency-dependent lower turning point)} and $\tau_\mathrm{A}=20$\,min (the acoustic pulse transit time across the cavity) as characteristic spatial and temporal scales to normalise the distance $r$ and time $t$ in Eq.~(\ref{eq:KG}), respectively. Thus, the acoustic cut-off frequency $\omega_\mathrm{ac}$ appears to be normalised by $\nu_0=1/\tau_\mathrm{A}\approx 833$\,$\mu$Hz.

\citet{1991ApJ...375L..35K} demonstrated that the governing Eq.~(\ref{eq:KG}) with the acoustic cut-off frequency $\omega_\mathrm{ac}(r)$ in form of Eq.~(\ref{eq:cutoff_profile}) and a harmonic source function $f(t,r)$ of fixed frequency $\omega$, localised at a height $r_0$ above the lower turning point within the subphotospheric cavity, has explicit analytical solution for the acoustic wave function $\psi(t,r)$. The Fourier power spectrum of this solution takes the form,
\begin{equation}\label{eq:spectrum}
    \mathcal{F}(\omega) = \left| \frac{\sin{\omega r_0}}{\omega\cos{\omega}+\sqrt{\alpha^2 - \omega^2}\sin{\omega}} \right|^2,
\end{equation}
where the source amplitude is set to unity due to the linear nature of Eq.~(\ref{eq:KG}), thus leaving the photospheric cut-off frequency $\alpha$ and the source position $r_0$ as the only free parameters in the problem. 

\begin{figure}
    \centering
    \includegraphics[width=\columnwidth]{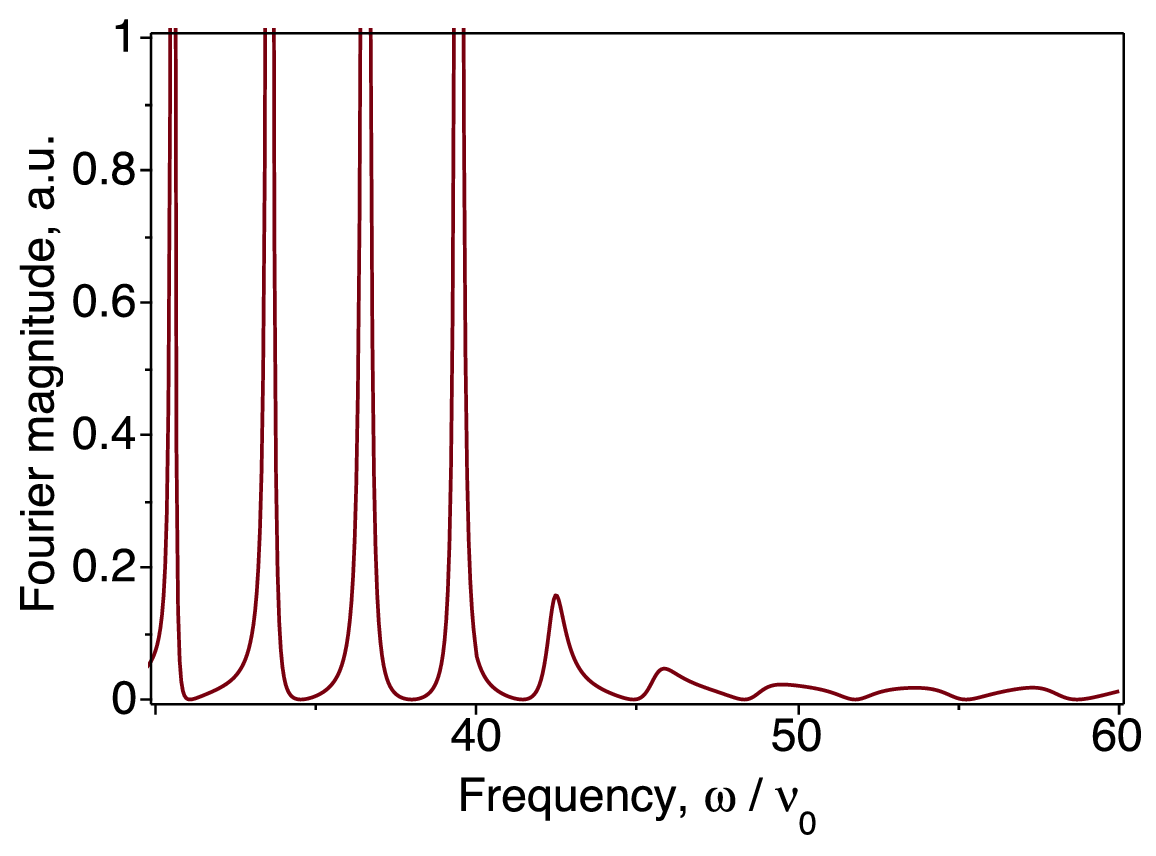}
    \caption{The Fourier power spectrum of low-frequency (below the acoustic cut-off frequency $\omega<\alpha$) and high-frequency ($\omega>\alpha$) acoustic waves in a subphotospheric cavity shown in Fig.~\ref{fig:cutoff_profile}. The spectrum is given by Eq.~(\ref{eq:spectrum}) with the acoustic cut-off frequency $\alpha/\nu_0 = 40$ and the wave source height $r_0/a=0.91$. The normalisation parameters $\nu_0$ and $a$ are the same as in Fig.~\ref{fig:cutoff_profile}.
    }
    \label{fig:spec_full}
\end{figure}

Figure~\ref{fig:spec_full} illustrates the spectrum given by Eq.~(\ref{eq:spectrum}) for $\alpha=40$ {(which is about 5300\,$\mu$Hz accounting for normalisation to $\nu_0=833\,\mu$Hz, $\alpha\nu_0/2\pi$)} and $r_0=0.91$ \citep[giving $(1-r_0)a \approx 6000$\,km beneath the photosphere for $a=0.1R_\odot$, cf.][]{1999ApJ...519..396K}.
In \citet{2024MNRAS.533.3387K}, the behaviour of the acoustic modes with $\omega<\alpha$, trapped within the acoustic cavity, with the cut-off parameter $\alpha$ was considered. In this work, we extend this analysis and study the effect of both $\alpha$ and $r_0$ on acoustic waves with $\omega>\alpha$. In this regime, acoustic waves can propagate through and form pseudomodes. Using the frequencies $\omega$ at which the Fourier power given by Eq.~(\ref{eq:spectrum}) {exhibits} peaks (Fig.~\ref{fig:spec_full}), Fig.~\ref{fig:cutoff_profile} illustrates examples of the spatial structure of acoustic modes trapped within the cavity (with $\omega<\alpha$) and a propagating pseudomode (with $\omega>\alpha$). These examples are obtained as numerical solutions of Eq.~(\ref{eq:KG}) with the \emph{pdsolve} function in the mathematical environment Maple 2024.2. The effective change in the pseudomode's wavelength when it escapes the cavity is connected with the effect of Klein-Gordon dispersion, i.e. the modification of the acoustic wave's propagation speed from $c_\mathrm{s}$ (inside the cavity with $\omega_\mathrm{ac}=0$) to $c_\mathrm{s}\omega/\sqrt{\omega^2-\alpha^2}$ (outside the cavity with $\omega_\mathrm{ac}=\alpha$); resulting in the effective wavelength increase by $\omega/\sqrt{\omega^2-\alpha^2}$. For example, for $\alpha=40$ and pseudomode frequency $\omega=54$ (both normalised to $\nu_0=833$\,$\mu$Hz, see Fig.~\ref{fig:cutoff_profile}), the pseudomode wavelength increases by about 1.5.

\begin{figure*}
    \centering
    \includegraphics[width=0.7\textwidth]{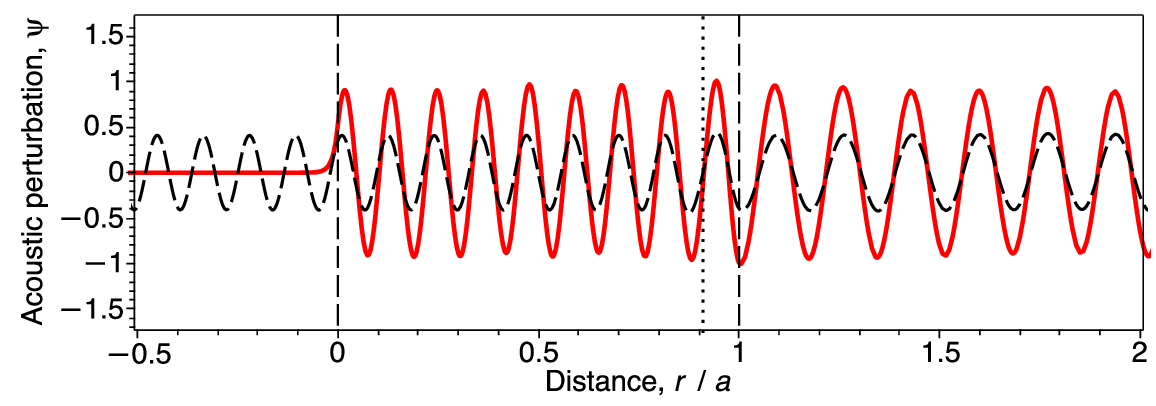}
    \includegraphics[width=0.7\textwidth]{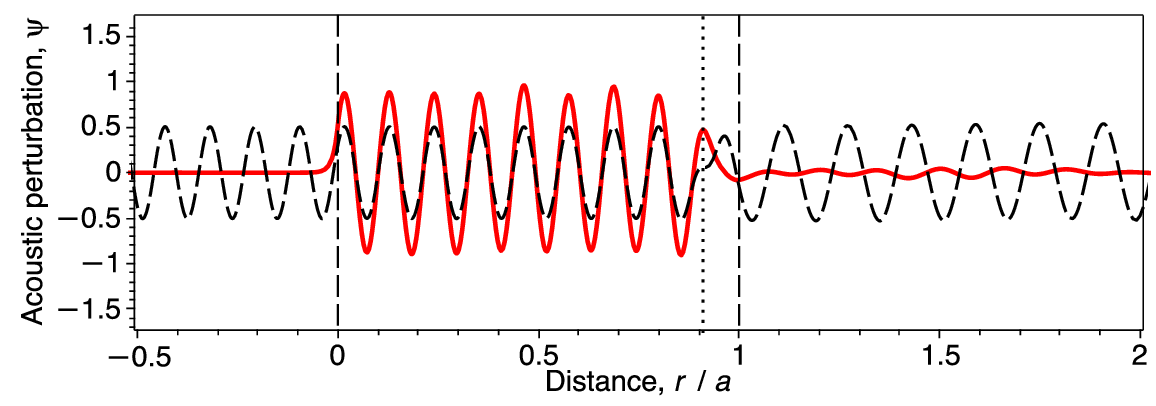}
    \caption{The snapshots of the high-frequency (above acoustic cut-off, $\omega>\alpha$) acoustic perturbation function $\psi(r)$ taken at $t=4.62$ (in solid red), obtained numerically from the governing Eq.~(\ref{eq:KG}) with the acoustic cut-off radial profile given by Eq.~(\ref{eq:cutoff_profile}), driver given by Eq.~(\ref{eq:driver}), and $\alpha/\nu_0=40$ and $r_0/a=0.91$.
    The top and bottom panels show the corresponding wave solutions in the regime of constructive (top) and destructive (bottom) interference with $\omega/\nu_0 = 53.65 + \pi/4r_0$ and $\omega/\nu_0 = 55.25 + \pi/4r_0$ (cf. Fig.~\ref{fig:spec_full}). The constant offset $\pi/4r_0$ is added to account for the initial phase shift caused by a dipole spatial structure of the driver Eq.~(\ref{eq:driver}) being an odd function of $r$.
    The vertical dashed and dotted lines show the lower ($r=0$) and upper ($r/a=1$) boundaries of the cavity and the {position} of the wave source ($r=r_0$), respectively.
    The dashed oscillatory signals in both panels illustrate the corresponding solutions with no reflection at the lower boundary, i.e. no interference effect.
    The normalisation parameters  $\nu_0$ and $a$ are the same as in Fig.~\ref{fig:cutoff_profile}.
    }
    \label{fig:snapshots}
\end{figure*}

In the considered acoustic wave model given by Eqs.~(\ref{eq:KG})--(\ref{eq:cutoff_profile}), pseudomode peaks in the high-frequency ($\omega>\alpha$) part of the Fourier spectrum (Fig.~\ref{fig:spec_full}) appear due to the effect of constructive or destructive interference of the acoustic wave reflected from the lower boundary of the acoustic cavity at $r=0$ with a newly emitted wave from the continuously operating source at $r=r_0$. Depending on the wave frequency $\omega$ and the source position $r_0$, the two waves may interfere constructively (i.e. adding up to produce a higher-amplitude signal and a local enhancement of the Fourier power) or destructively (i.e. cancelling each other and leading to a local suppression of the Fourier power). To illustrate this, we solve Eq.~(\ref{eq:KG}) numerically with the \emph{pdsolve} function in Maple 2024.2, with the following driver,
\begin{equation}\label{eq:driver}
    f(t,r) = (r-r_0)\exp\left[ -\frac{(r-r_0)^2}{\Delta r}\right]\sin(\omega t),
\end{equation}
where we set $r_0=0.91$ (as used for the spectrum shown in Fig.~\ref{fig:spec_full} for direct comparison), $\Delta r = 0.001$ (to keep the wave source local) and choose $\omega = 53.65 + \pi/4r_0$ and $\omega = 55.25 + \pi/4r_0$ which correspond to the peak and trough in the Fourier spectrum shown in Fig.~\ref{fig:spec_full}. {A value of $\Delta r = 0.001$ corresponds to around 60\,km, which is well below the upper limit of 550\,km set by \citet{1994ApJ...428..827K}.} The constant offset of  $\pi/4r_0$ is added to account for the phase shift caused by the dipole spatial structure of our driver given by Eq.~($\ref{eq:driver}$) vs. a monopole Dirac delta function of $(r-r_0)$ considered as a driver in \citet{1991ApJ...375L..35K}, the exact solution of which was used to derive the Fourier spectrum given by Eq.~(\ref{eq:spectrum}) shown in Fig.~\ref{fig:spec_full}. Thus, Fig.~\ref{fig:snapshots} shows the snapshots of our numerical solutions for the acoustic wave function $\psi(r)$. Here, for $r>r_0$ (and $r>1$), we observe the increase of the wave amplitude by approximately a factor of 2 (vs. the case with no reflection at the lower boundary of the cavity, i.e. no interference) in the regime of constructive interference and the decrease of the wave amplitude almost to zero in the regime of destructive interference.

\begin{figure*}
    \centering
    \includegraphics[width=0.8\textwidth]{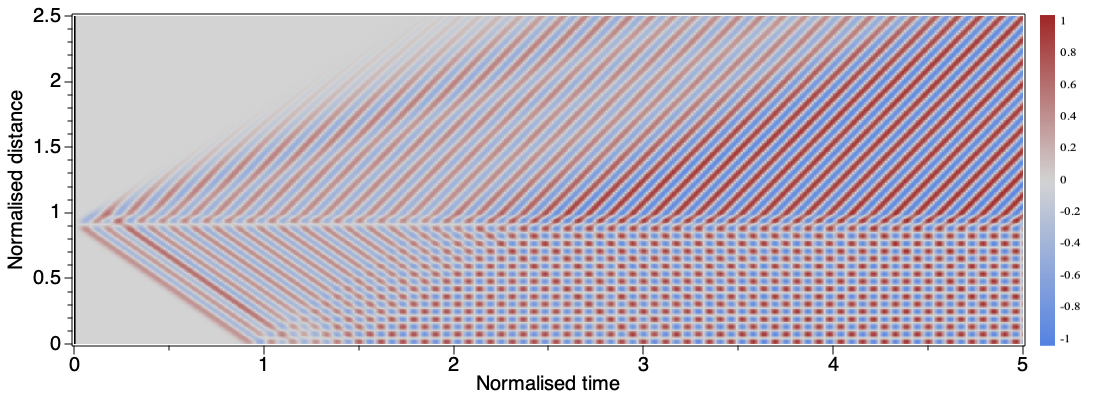}
    \includegraphics[width=0.8\textwidth]{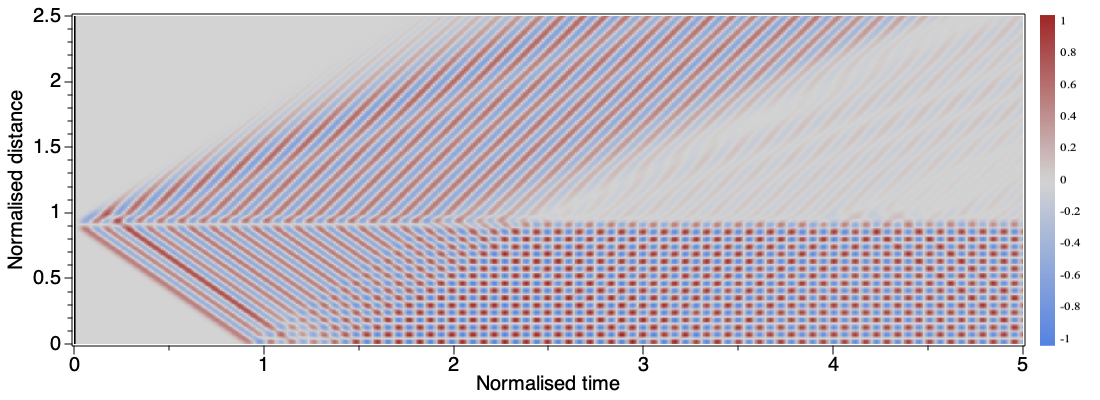}
    \caption{Time-distance plots illustrating the full spatio-temporal evolution of the high-frequency acoustic perturbation function $\psi(t,r)$ obtained numerically from Eq.~(\ref{eq:KG}) and the same model setup as described in the caption to Fig.~\ref{fig:snapshots}. The top and bottom panels illustrate the regimes of constructive and destructive intereference of acoustic waves, forming pseudomode peaks and troughs in the high-frequency part ($\omega>\alpha$) of the Fourier spectrum shown in Fig.~\ref{fig:spec_full}. In both panels, the colour bars show $\psi(t,r)$ normalised to its maximum; the time and distance are normalised to $\tau_\mathrm{A}=20$\,min and $a=0.1 R_\odot$.
    }
    \label{fig:td_maps}
\end{figure*}

The full spatio-temporal evolution of $\psi(t,r)$ in both constructive and destructive regimes is displayed in Fig.~\ref{fig:td_maps} as time–distance plots. The wave source (Eq.~(\ref{eq:driver})), located at $r_0 = 0.91$, excites high-frequency ($\omega > \alpha$) acoustic waves that propagate in both directions: upwards, towards the higher solar atmosphere, and downwards, towards the solar interior. As discussed above, the high-frequency upward-propagating wave freely escapes the cavity, whereas the downward-propagating wave reflects at $r = 0$ (at the lower boundary of the cavity with an infinite potential barrier). In addition to the enhancement/suppression of the signal when the reflected wave interferes with the upward-propagating wave at $r>r_0$ {(see the absence of power in the right-hand triangle region in the lower panel but the enhanced power in the upper panel)}, these plots show the formation of the standing wave pattern inside the cavity (at $r<r_0$) due to the interaction of the downward-propagating wave with the reflected wave. From this perspective, the subphotospheric acoustic cavity acts as an effective Fabry-P\'erot interferometer, supporting standing-mode resonances inside the cavity and producing constructive or destructive interferometric features outside it that result in a discrete spectrum of high-frequency acoustic waves.

\section{Effects of cut-off and source position  on pseudomode frequency spectrum}
\label{sec:alpha_r0_effects}

As shown in Fig.~\ref{fig:spec_full}, the Fourier spectrum described by Eq.~(\ref{eq:spectrum}) captures both the low-frequency ($\omega < \alpha$) acoustic waves confined within the cavity and the high-frequency ($\omega > \alpha$) propagating waves that form pseudomodes. For $\omega > \alpha$, the denominator in Eq.~(\ref{eq:spectrum}) becomes complex-valued. By separating this term into its real and imaginary parts and then computing the magnitude, we can derive an explicit dependence of the Fourier power of pseudomode oscillations, $\mathcal{F}_\mathrm{pseudo}$ upon frequency,
\begin{equation}\label{eq:spec_pseudo}
  \mathcal{F}_\mathrm{pseudo}=\frac{\sin(\omega r_0)^2}{\omega^2-\alpha^2\sin(\omega)^2}.
\end{equation}

\begin{figure*}
    \centering
    \includegraphics[width=0.4\textwidth]{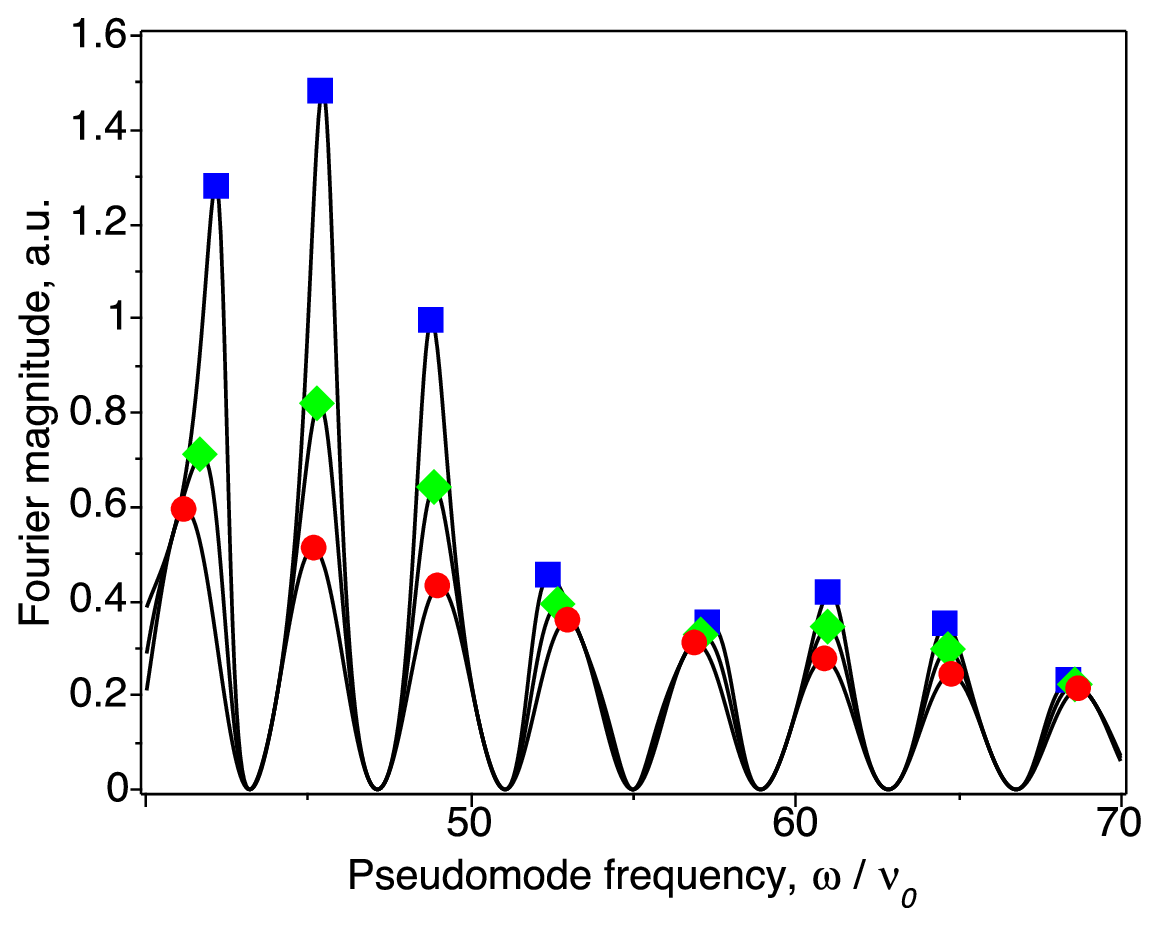}
    \includegraphics[width=0.4\textwidth]{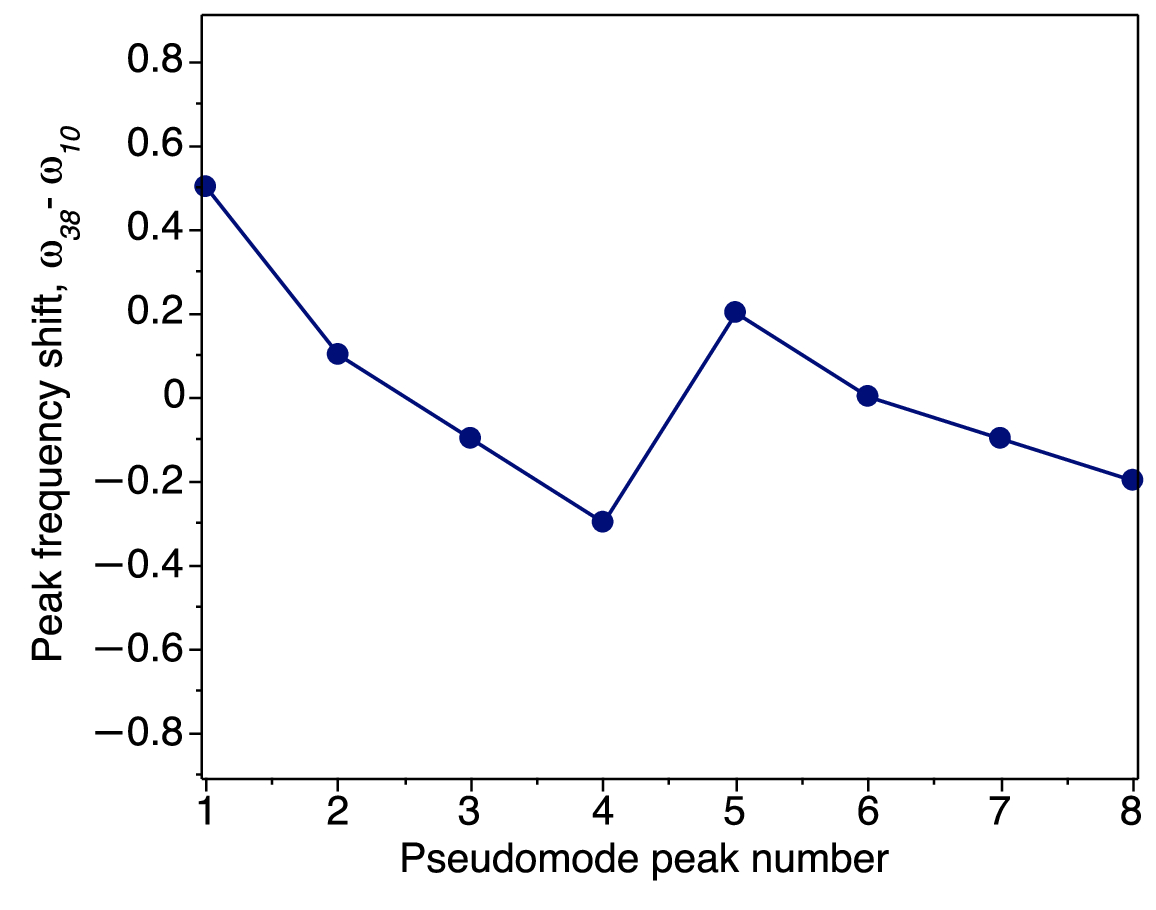}
    \includegraphics[width=0.4\textwidth]{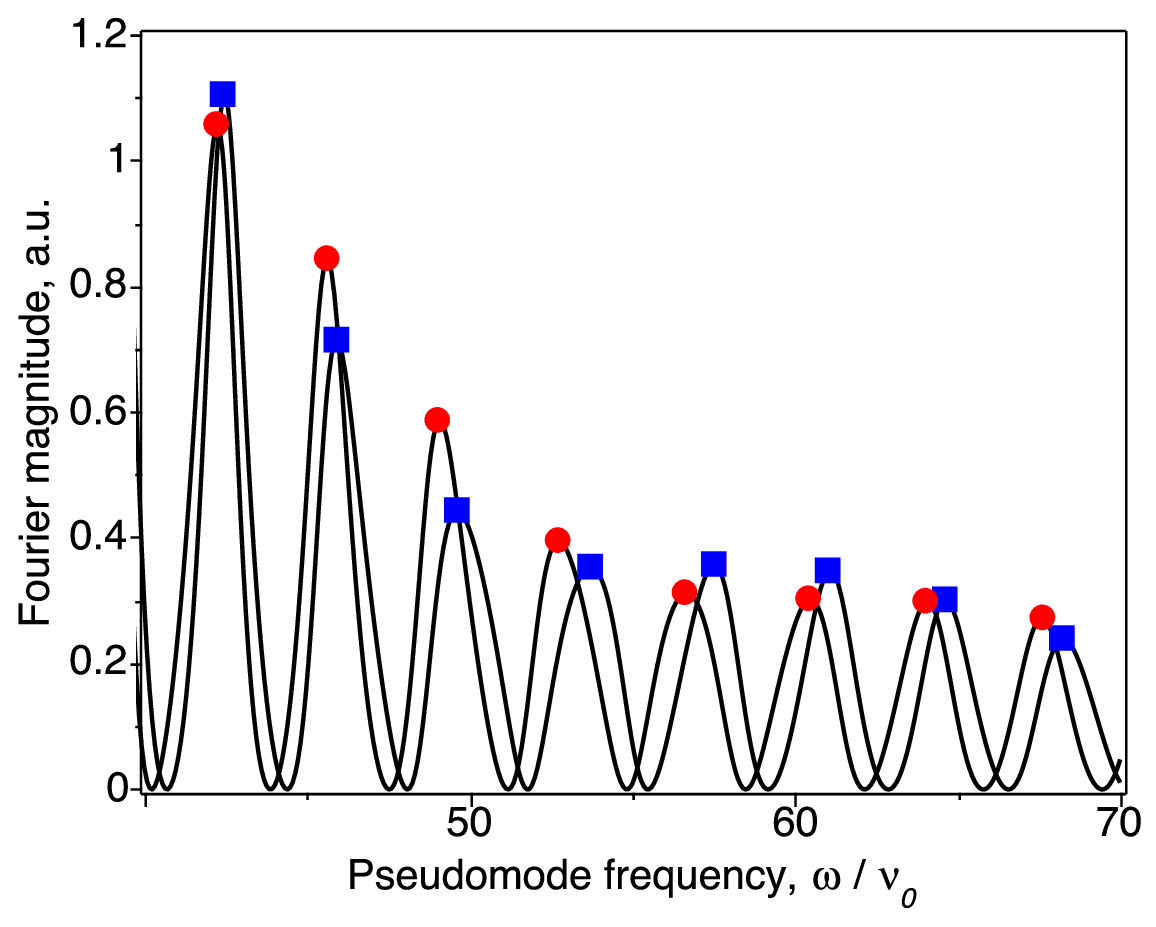}
    \includegraphics[width=0.4\textwidth]{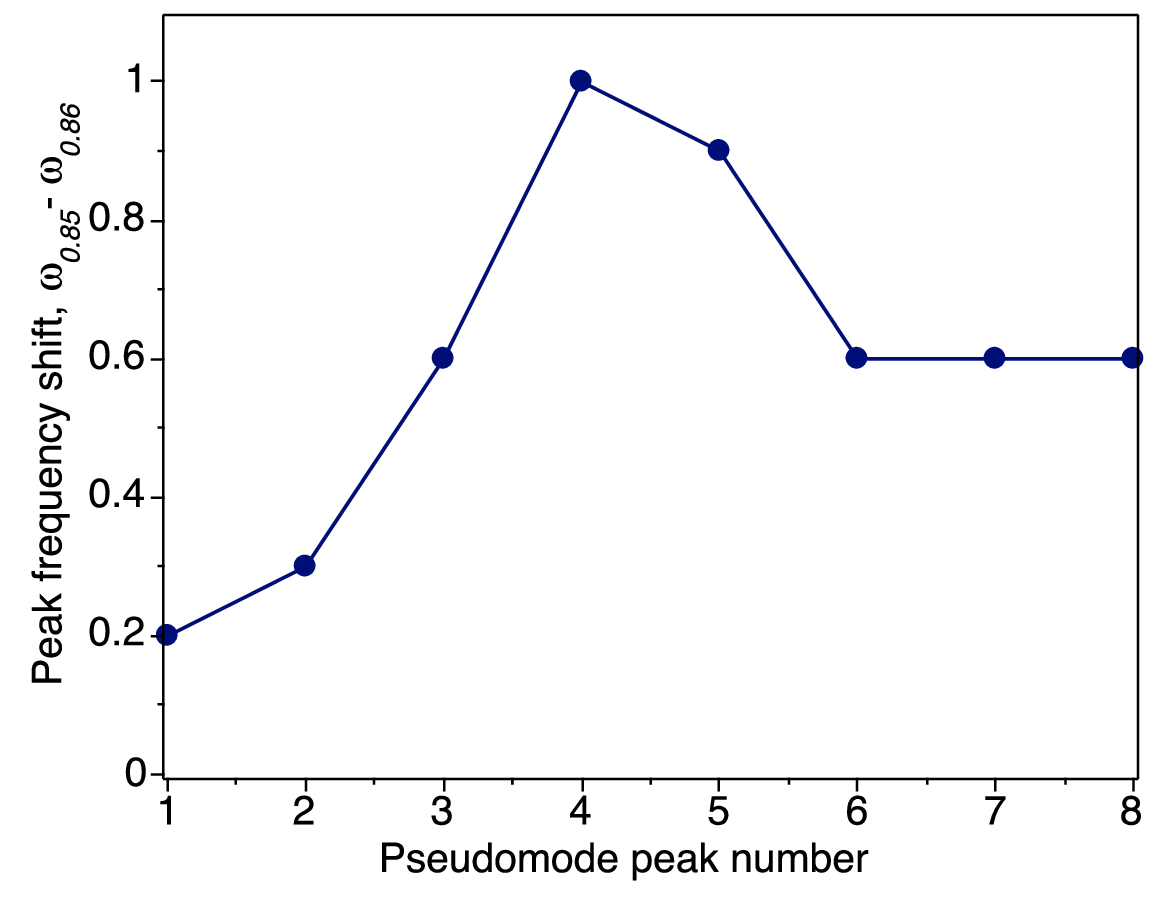}
    \caption{Left column: pseudomode oscillation spectra given by Eq.~(\ref{eq:spec_pseudo}) for the acoustic cut-off frequency $\alpha = 38$ (blue boxes),  30 (green diamonds), 10 (red circles) and fixed source height $r_0=0.8$ (top); and the source height $r_0 = 0.85$ (blue boxes), 0.86 (red circles) and fixed cut-off frequency $\alpha=30$ (bottom).
    Right column: the corresponding pseudomode peak frequency shifts caused by the effect of $\alpha$ varying from 10 to 38 (top) and $r_0$ varying from 0.86 to 0.85 (bottom). The normalisation of all parameters is the same as in Fig.~\ref{fig:cutoff_profile}.
    }
    \label{fig:spec_pseudo}
\end{figure*}

\noindent Figure~\ref{fig:spec_pseudo} illustrates $\mathcal{F}_\mathrm{pseudo} (\omega)$ given by Eq.~(\ref{eq:spec_pseudo}) and its dependence on the model parameters $\alpha$ and $r_0$. When the source height $r_0$ is kept fixed, the positions of the pseudomode spectral peaks are found to vary with the photospheric cut-off frequency $\alpha$ (see the top row panels in Fig.~\ref{fig:spec_pseudo}). Moreover, the character of this dependence differs {(both in magnitude and in sign)} across frequency bands. For example, the frequencies of the pseudomode peaks at $40 < \omega < 45$ and $55 < \omega < 60$ are seen to increase with $\alpha$, whereas the peaks at $50 < \omega < 55$ and $65 < \omega < 70$ decrease with $\alpha$ (see the top left panel in Fig.~\ref{fig:spec_pseudo}). Likewise, the pseudomode peaks at $45 < \omega < 50$ and $60 < \omega < 65$ seem to be less sensitive to $\alpha$. The top right panel of Fig.~\ref{fig:spec_pseudo} illustrates how such a non-monotonic dependence of the $\mathcal{F}_\mathrm{pseudo}$ peak frequencies on $\alpha$ can be captured in practice. Taking the pseudomode peak positions for $\alpha = 10$ as a reference, one can compute the differences between these and the corresponding peak positions for another value of $\alpha$, e.g. $\alpha = 38$. As shown in the top right panel, these frequency shifts can be either positive or negative depending on the pseudomode peak number, indicating that the frequencies of the $\mathcal{F}_\mathrm{pseudo}$ peaks may increase or decrease with $\alpha$. Analogously, we study the behaviour of the positions of the pseudomode $\mathcal{F}_\mathrm{pseudo}$ spectral peaks with the source height $r_0$, keeping the cut-off parameter $\alpha$ fixed (the bottom row of Fig.~\ref{fig:spec_pseudo}). In this case, the frequencies of all pseudomode peaks are found to decrease with $r_0$ (bottom left panel), which is reflected by the same sign of the peak frequency shift caused by $r_0$ for all pseudomode peak numbers (bottom right panel).

\begin{figure*}
    \centering
    \includegraphics[width=0.4\textwidth]{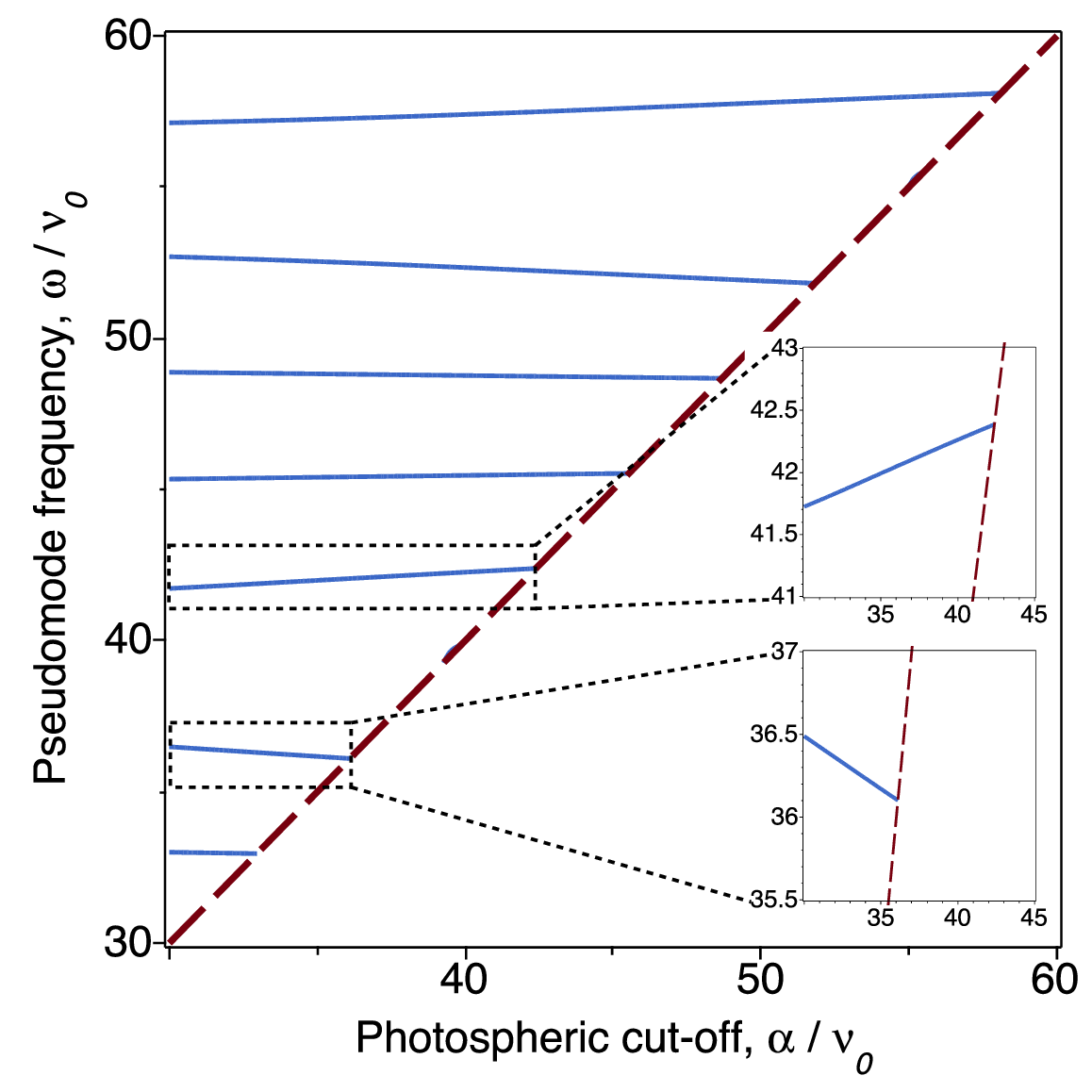}
    \includegraphics[width=0.4\textwidth]{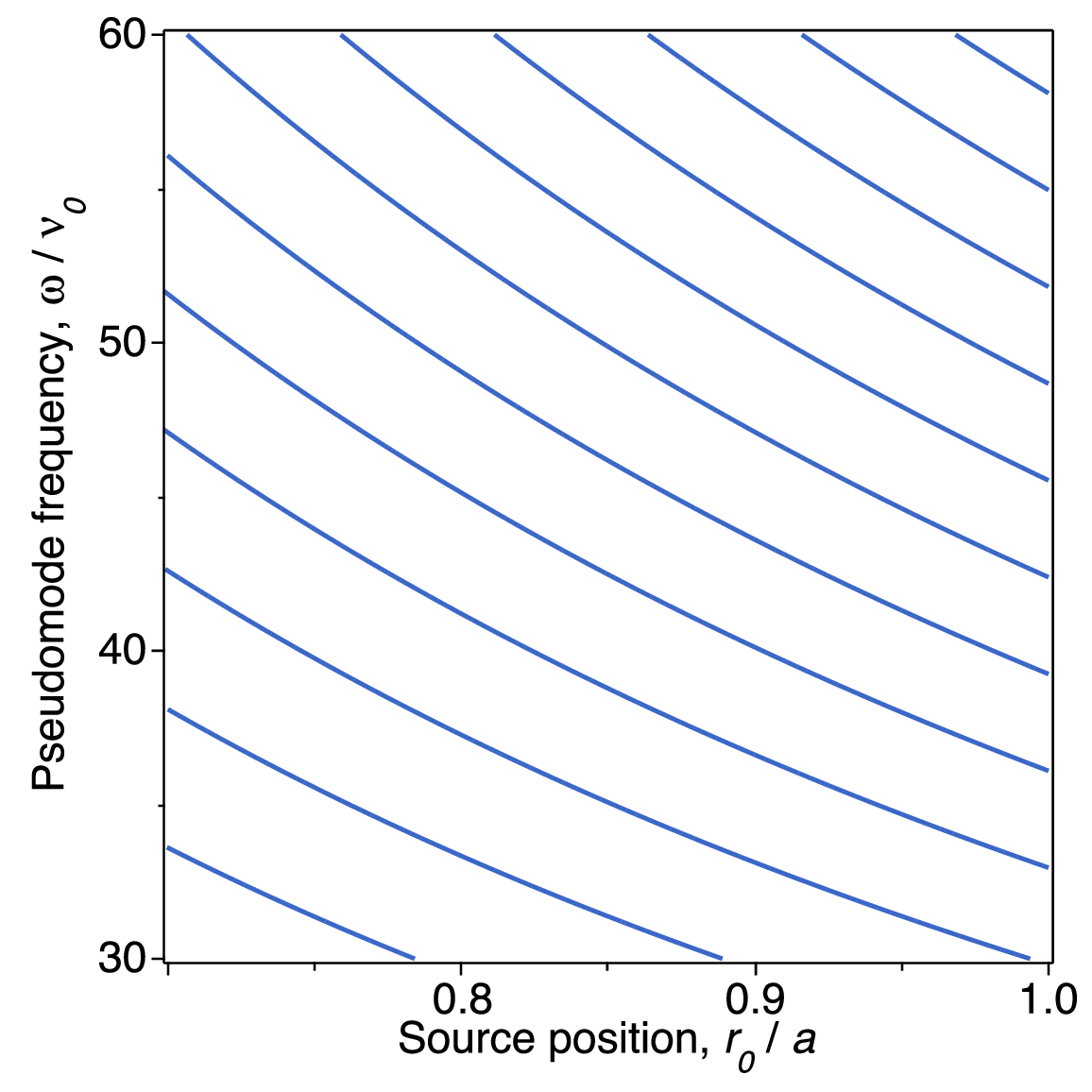}
    \caption{Left: effect of acoustic cut-off $\alpha$ with fixed source height $r_0=0.8$ on the frequency of several pseudomode radial harmonics with $\omega>\alpha$. The burgundy dashed line indicate $\omega=\alpha$.
            Right: effect of the source height $r_0$ with cut-off frequency $\alpha \to 0$ on the frequency of several pseudomode radial harmonics.
            Both are obtained as numerical solutions of Eq.~(\ref{eq:disp_pseudo}). The normalisation of all parameters is the same as in Fig.~\ref{fig:cutoff_profile}.
    }
    \label{fig:disp}
\end{figure*}

The exact equation connecting the pseudomode frequency peaks with the cut-off frequency $\alpha$ and the source height $r_0$ can be obtained from Eq.~(\ref{eq:spec_pseudo}) using $d\mathcal{F}_\mathrm{pseudo}/d\omega = 0$ and excluding the solutions of $\sin(\omega r_0)=0$ which stand for troughs in the pseudomode spectrum,
\begin{equation}\label{eq:disp_pseudo}
    \frac{\tan(\omega r_0)}{r_0}+\frac{2(\omega^2-\alpha^2\sin(\omega)^2)}{\alpha^2\sin(2\omega)-2\omega}=0.
\end{equation}
Thus, in Eq.~(\ref{eq:disp_pseudo}), we managed to isolate the effects of $r_0$ and $\alpha$ (i.e. mode-related and medium-related effects) on the pseudomode peak frequencies in the first and second terms, respectively. It can be treated as an effective dispersion relation of pseudomode oscillations, describing how pseudomode peak frequencies get modified by the wave source position and the photospheric cut-off effect \citep[cf. analogous Eq.~(6) for trapped p-modes in][which is shown to have no dependence upon $r_0$]{2024MNRAS.533.3387K}. Figure~\ref{fig:disp} shows numerical solution of Eq.~(\ref{eq:disp_pseudo}) for pseudomode peak frequencies $\omega(\alpha)$ and $\omega(r_0)$ (for fixed $r_0$ and $\alpha$, respectively) for several radial harmonics. Similarly to Fig.~\ref{fig:spec_pseudo}, here we observe that the gradient of $\omega(\alpha)$ (left panel) differs across the pseudomode harmonics, so that the pseudomode frequency may both increase and decrease with $\alpha$ (see e.g. the harmonics with $\omega$ around 42 and 36, respectively). In contrast, the gradient of $\omega(r_0)$ (right panel) is found to be negative for all pseudomode harmonics, indicating the decrease of pseudomode peak frequency with $r_0$. We note that for $\omega\gg\alpha$, Eq.~(\ref{eq:disp_pseudo}) further simplifies to $\tan(\omega r_0)=\omega r_0$ with an approximate solution $\omega \approx (n+1/2)\pi/r_0$, explicitly demonstrating the inverse proportionality between pseudomode peak frequency $\omega$ and $r_0$.

\section{Comparison with observations}
\label{sec:observ}

\subsection{Best-fitting the observed pseudomode spectrum}
\label{sec:best-fit}

\begin{figure}
    \centering
    \includegraphics[width=\columnwidth]{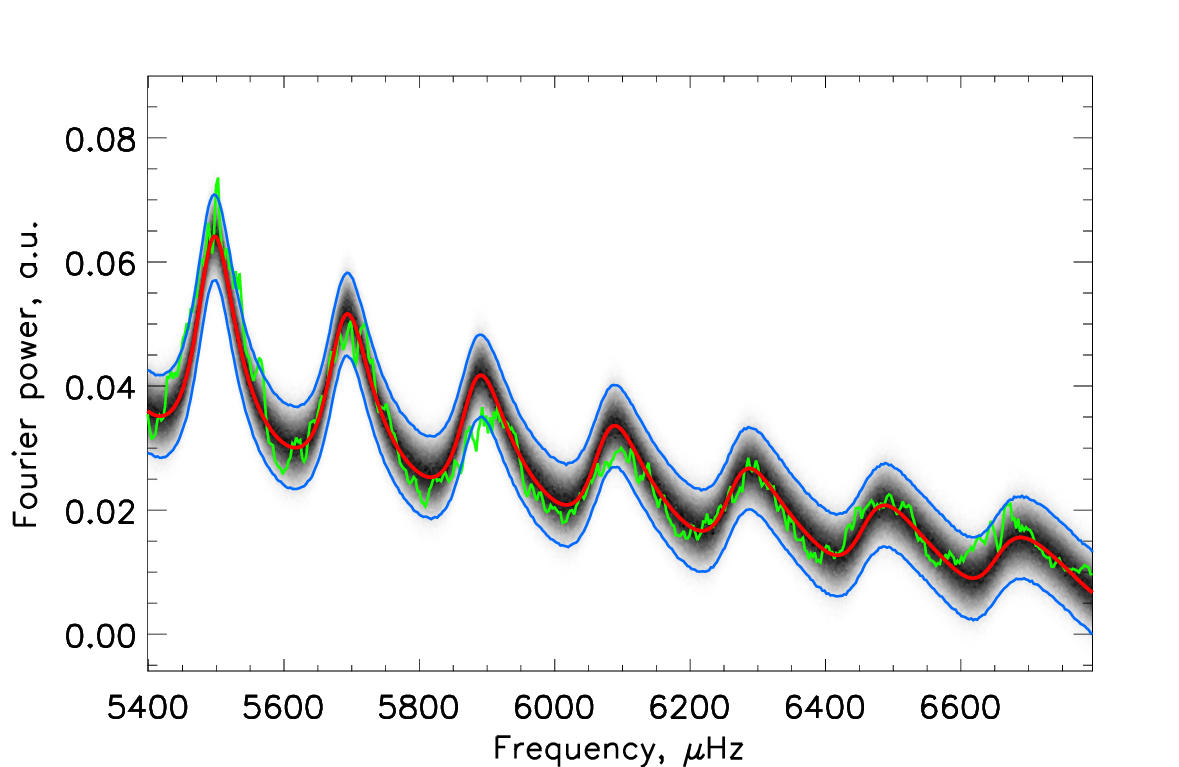}
    \caption{Observed pseudomode spectrum (green) obtained from the GONG data for the $\ell = 100, m = 0$ mode, best-fitted by Eq.~(\ref{eq:fit_model}) with Bayesian MCMC approach (red), as described in Sec.~\ref{sec:best-fit}. The 95\% credible intervals and posterior predictive distribution are shown in blue and in shades of grey, respectively.
    }
    \label{fig:best-fit}
\end{figure}

To validate the theoretically derived pseudomode spectrum given by Eq.~(\ref{eq:spec_pseudo}), we best-fit it to the observed spectrum (Fig.~\ref{fig:best-fit}, green line) obtained using data from the Global Oscillation Network Group (GONG). For the $\ell = 100, m = 0$ mode, the $\sim$22-year time series was divided into consecutive subseries. A Lomb-Scargle periodogram was computed for each subseries with a high duty cycle, and the resulting periodograms were averaged to produce a single high-SNR spectrum. This spectrum was then smoothed using a 10~$\mu$Hz boxcar window, and cropped between 5400--6800~$\mu$Hz.

By accounting for the frequency normalisation to $\nu_0$ in Eq.~(\ref{eq:spec_pseudo}), our best-fitting function takes the form,
\begin{equation}\label{eq:fit_model}
    \mathcal{F}_\mathrm{pseudo}=A\frac{\sin(2\pi r_0\nu/\nu_0)^2}{(2\pi\nu/\nu_0)^2-\alpha^2\sin(2\pi\nu/\nu_0)^2} + \frac{B}{(2\pi\nu/\nu_0)^C} + D,
\end{equation}
where $\nu$ is the observed frequency, and the additional parameters $A$, $B$, $C$, and $D$ account for the overall scaling of the observed spectrum, the power-law background noise, and the vertical offset. Thus, together with $\alpha$, $r_0$, and $\nu_0$, the model function given by Eq.~(\ref{eq:fit_model}) contains seven free parameters to be determined by best-fitting.

We use Bayesian inference with Markov chain Monte Carlo
(MCMC) sampling approach, implemented by the open-access Solar Bayesian Analysis Toolkit \citep{2021ApJS..252...11A} to best-fit the observed pseudomode spectrum by the model given by Eq.~(\ref{eq:fit_model}). The resulting best-fitting curve and the posterior predictive distribution (i.e. where new data are expected to appear according to the model and the existing observations) are shown in Fig.~\ref{fig:best-fit} (in red and in shades of grey, respectively), with the corresponding 95\% credible intervals (in blue).

As seen in Fig.~\ref{fig:best-fit}, the pseudomode model spectrum given by Eq.~(\ref{eq:fit_model}) successfully reproduces the observed spectrum for reasonable values of the model parameters, which are summarised in Table~\ref{tab:params}. In particular, the MCMC-inferred values of $\nu_0 \approx 385$\,$\mu$Hz and $\alpha\approx75.9$ yield the acoustic cut-off frequency $\nu_\mathrm{ac} = \alpha\nu_0/2\pi\approx 4656$\,$\mu$Hz and a cavity transit time $\tau_\mathrm{A}=\nu_0^{-1}\approx 43$\,min. Combined with the radial dependence of the sound speed \citep[e.g. via the standard solar model,][]{2021LRSP...18....2C}, $c_\mathrm{s}^2= (\gamma - 1)GM_\odot(r^{-1}-R_\odot^{-1})$, it allows us to estimate the cavity depth, $a$ as $\tau_\mathrm{A}=\int_{R_\odot-a}^{R_\odot}dr/c_\mathrm{s}$, which gives $a\approx 0.5R_\odot$. In general, greater penetration depths are expected for high-frequency ($\nu>\nu_\mathrm{ac}$), $\ell=100$--150 modes. For example, using the standard Lamb frequency dependence upon radial distance and spherical degree, $S_\ell = \sqrt{\ell(\ell+1)}c_\mathrm{s}(r)/r$, one obtains $a=0.1$--0.4$R_\odot$ for $S_\ell/2\pi=5000$--10000\,$\mu$Hz and $\ell=100$--150. Thus, the obtained value of $a=0.5R_\odot$ appears to be overestimated. The inferred value of $r_0\approx0.962$ gives $(1-r_0)a\approx 13$\,Mm below the surface
for the wave source location. It appears beyond the commonly accepted interval for p mode and pseudomode source location \citep[100--1000\,km, see e.g.][]{1994ApJ...428..827K, 1999MNRAS.309..761C}, with the mismatch likely caused by an overestimation of the cavity depth $a$.

\begin{table*}
\centering
\renewcommand{\arraystretch}{1.35} 
\caption{Best-fitting parameters of the pseudomode model spectrum given by Eq.~(\ref{eq:fit_model}), used to approximate the observed spectrum as shown in Fig.~\ref{fig:best-fit}.}
\begin{tabular}{lcl}
\hline
Parameter & Value & Comment \\
\hline
$\nu_0$ ($\mu$Hz) 
    & $385.5^{+0.3}_{-0.4}$ 
    & providing transit time $\tau_\mathrm{A} = \nu_0^{-1} = 43.23^{+0.04}_{-0.03}$ min \\

$\alpha$ 
    & $75.9^{+2.3}_{-3.0}$ 
    & normalised to $\nu_0$, so that $\nu_\mathrm{ac} = \dfrac{\alpha\nu_0}{2\pi} = 4656^{+138}_{-185}~\mu$Hz \\

$r_0$ 
    & $0.962^{+0.001}_{-0.002}$ 
    & normalised to cavity depth $a$ to be obtained from $\tau_\mathrm{A}$ and $c_\mathrm{s}(r)$ \\

$A$ 
    & $75.1^{+12.7}_{-9.5}$ 
    & --- \\

$B$ 
    & $10.0^{+0.02}_{-4.3}$ 
    & --- \\

$C$ 
    & $0.93^{+0.003}_{-0.17}$ 
    & --- \\

$D$ 
    & $-0.12^{+0.004}_{-0.03}$ 
    & --- \\
\hline
\end{tabular}
\label{tab:params}
\end{table*}

\subsection{11-year pseudomode frequency shifts}

\begin{figure}
    \centering
    \includegraphics[width=\columnwidth]{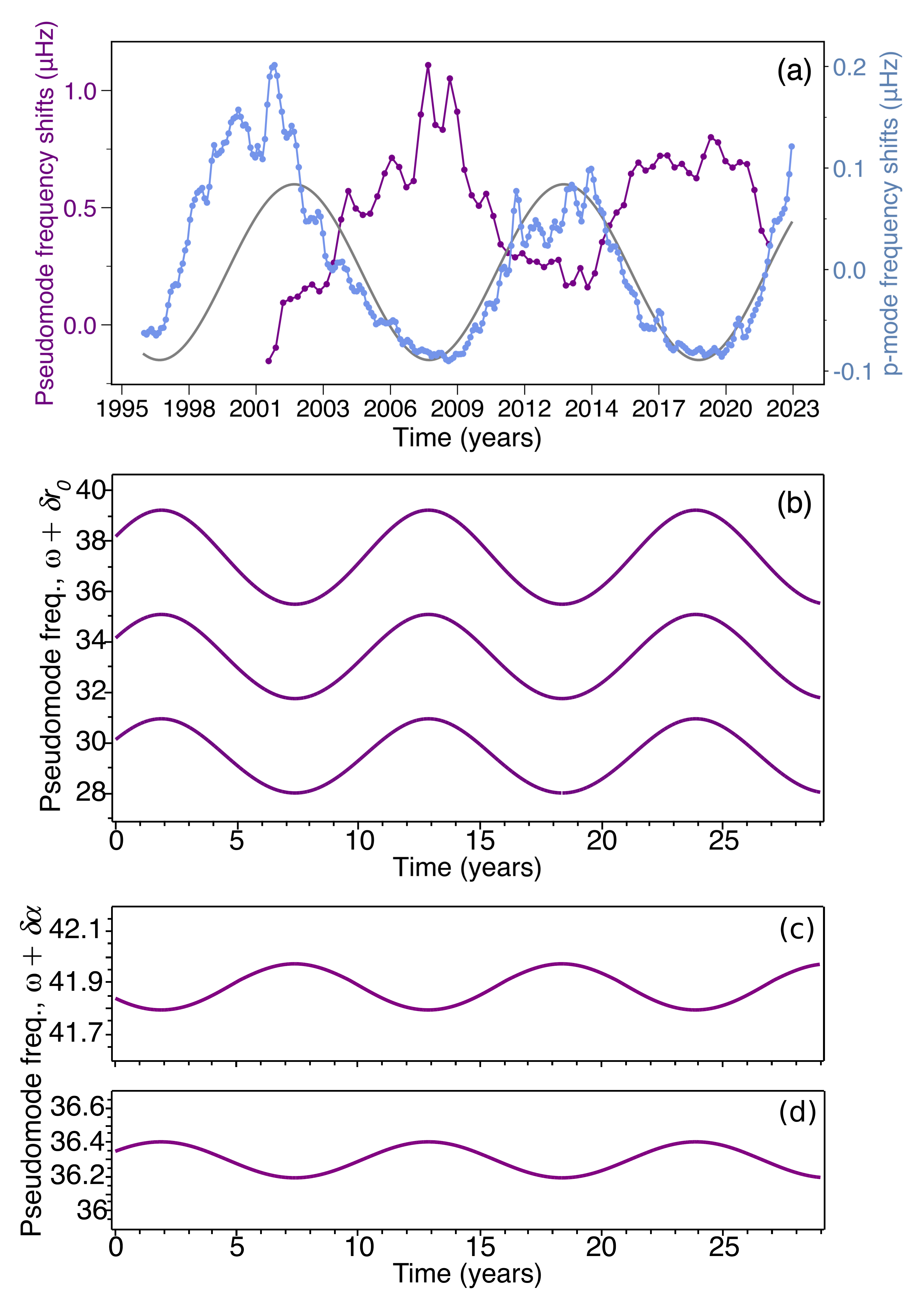}
    \caption{(a) Solar pseudomode (5600-6800~$\mu$Hz, purple markers) and p-mode (1900-4100~$\mu$Hz, blue markers) frequency shifts averaged over all azimuthal orders for harmonic degrees 0~$\leq$~$\textit{l}$~$\leq$~150. The error bars are too small to be seen. The frequency shifts have been linearly re-scaled and shifted for the purpose of this comparison. A sine wave (grey line) with a period of 11 years has been superimposed on the p-mode shifts.
    (b) Numerical solution of Eq.~(\ref{eq:disp_pseudo}) for the pseudomode frequency $\omega$ with the acoustic cut-off $\alpha\to 0$ and the source position $r_0 + \delta r_0$ modulated by a harmonic signal shown in grey in panel (a) with 5\% amplitude.
    (c)-(d) Numerical solution of Eq.~(\ref{eq:disp_pseudo}) for the pseudomode frequency $\omega$ with the source position $r_0 = 0.8$ and the cut-off frequency $\alpha + \delta\alpha$ modulated by a harmonic signal shown in grey in panel (a) with 5\% amplitude.
    Different curves in panels (b)-(d) correspond to different radial pseudomode harmonics.
    The pseudomode frequency in panels (b)-(d) is normalised to $\nu_0=833$\,$\mu$Hz.
    }
    \label{fig:shifts}
\end{figure}



We now apply the links of the pseudomode peak frequency with the photopsheric acoustic cut-off parameter $\omega(\alpha)$ and the wave source postion $\omega(r_0)$, derived in Sec.~\ref{sec:alpha_r0_effects} and illustrated in Fig.~\ref{fig:disp}, to the observed phase behaviour of solar pseudomodes with the 11-year magnetic activity cycle.

Indeed, the frequencies of the pseudomodes, like those of the p~modes, have been observed to vary with time throughout the 11-year sunspot cycle \citep{hathaway2015solar}. The observed relationship between p-mode and pseudomode frequency shifts is illustrated in Fig.~\ref{fig:shifts}(a) where p-mode shifts (blue markers) are shown alongside pseudomode shifts (purple markers), with both being measured using data from GONG. Solar p-modes frequency shifts were calculated in the range 1900-4100~$\mu$Hz and pseudomodes between 5600-6800~$\mu$Hz (i.e. above the acoustic cut-off frequency which is around 5100~$\mu$Hz), and both display the average of all azimuthal orders, $\textit{m}$, across harmonic degrees 0~$\leq$~$\textit{l}$~$\leq$~150. Pseudomode frequency shifts and their uncertainties were generated using a resampled periodogram cross-correlation method (see \citealt{millson2024latitudinal} for details) using GONG data. The p-mode frequency shifts were determined using the GONG fitted mode frequencies from the mfv1f data files supplied on their website. These provide frequencies for each of the $n$, $\ell$ and $m$ components. When determining the frequency shifts, we only used modes for which successful fits to the data were obtained for each of the GONG 108\,d subsections across the entire epoch considered. The frequency shifts were measured in the manner described by \citet{2022MNRAS.515.2415M}, and summarised here. Frequency shifts for each individual mode were obtained by finding the difference between the mode frequency in an individual 108\,d segment and the weighted average frequency of that mode across all segments. The weighted mean of the frequency shift was then determined using these individual mode frequency shifts.
On visual inspection of Fig.~\ref{fig:shifts}(a), the anti-phase behaviour of the pseudomode frequency shifts with those of p~modes (and hence the 11-year cycle) is clearly evident. When p-mode frequencies increase in value as solar magnetic activity intensifies, the pseudomodes behave in opposition, decreasing in value to reach a minimum at solar maximum. Between cycle maximum and minimum, the opposite is also true: p-mode frequencies decrease towards solar minimum, whereas the pseudomode frequency variation reaches a maximum value.

We approximate the observed 11-year variation in the p-mode frequency shifts in Fig.~\ref{fig:shifts} by a harmonic signal $\propto -\sin(2\pi t/11 + 0.5)$ (grey line), with a fixed period of 11 years. This serves as a proxy for the 11-year solar cycle modulation in our problem. While the average sunspot cycle (between solar cycles 1 to 22) is almost exactly 11 years in length, it is worth noting that the duration of individual cycles can vary by a few years \citep{hathaway2015solar}. 
Furthermore, a simple sinusoid does not capture the fact that the amplitude of each cycle is known to vary.
Nevertheless, the sine-wave approximation described above is adequate and sufficient for the purposes of this study. 
Following the approach of \citet{2024MNRAS.533.3387K}, we use this 11-year harmonic signal to independently modulate the source position, $r_0 + \delta r_0$, and the cut-off frequency, $\alpha + \delta \alpha$, in Eq.~(\ref{eq:disp_pseudo}). We then examine whether the pseudomode frequency, $\omega(r_0+\delta r_0)$ and $\omega(\alpha+\delta \alpha)$ determined by Eq.~(\ref{eq:disp_pseudo}), responds in phase or out of phase with the applied 11-year harmonic modulation of the model parameters. The modulation amplitude was set to 5\% for both $r_0$ and $\alpha$. 

Figure~\ref{fig:shifts}(b) shows that the observed anti-phase behaviour of pseudomode frequency shifts with the 11-year cycle can be readily reproduced (at least qualitatively) by the modulation of the wave source position $r_0$. Indeed, in this case, due to the negative gradient of the $\omega(r_0)$ dependence (Fig.~\ref{fig:disp}, right panel), the frequencies of all radial pseudomode harmonics respond in anti-phase with the imposed 11-year cyclic modulation of $r_0$. In other words, when $r_0$ increases harmonically around some mean value, the corresponding pseudomode frequency decreases and vice versa. This scenario appears to be different for the modulation by the acoustic cut-off frequency $\alpha$, as shown in Fig.~\ref{fig:shifts}(c-d). In this case, the phase behaviour of the resulting pseudomode frequency modulation acquires dependence on the radial harmonic number. For example, for the pseudomode harmonic with $\omega$ being around 36 (panel (d)), the pseudomode frequency responds in anti-phase as seen in observations (cf. the negative gradient of $\omega (\alpha)$ for this mode in Fig.~\ref{fig:disp}, left panel). In contrast, for the same 11-year modulation of $\alpha + \delta \alpha$, the pseudomode harmonic with $\omega$ being around 42 (panel (c)) appears to behave in phase with the cycle (as suggested by the positive gradient of $\omega (\alpha)$ for this mode in Fig.~\ref{fig:disp}, left panel). We also note that in terms of our model, the apparent modulation depth of the pseudomode frequency via $\alpha$-mechanism (i.e. by 11-year cyclic variations of the acoustic cut-off frequency $\alpha$) is seen to be far lower than that in $r_0$-mechanism (by cyclic variations of the wave source position $r_0$), as evident from the visual inspection of the signals in panel (b) and panels (c-d) of Fig.~\ref{fig:shifts}.

\section{Discussion and Conclusions}
\label{sec:disc}

We studied how the subsurface excitation source location and the photospheric acoustic cut-off frequency influence the formation, frequency structure, and solar-cycle variability of helioseismic pseudomodes using an analytical Klein-Gordon cavity model supported by numerical simulations, and performed a comparison with GONG observations. By treating the subsurface acoustic cavity as an effective Fabry-P\'erot interferometer, we derived the corresponding dispersion relation that isolates the effects of the source depth and acoustic cut-off on the frequency peak-trough structure of pseudomodes. We further tested the model through Bayesian MCMC fitting of the observed pseudomode spectrum and by examining how cyclic variations in the source location and cut-off frequency reproduce the phase behaviour of the 11-year pseudomode frequency shifts. The main findings of this work can be summarised as follows:

\begin{itemize}
    \item Through analytical and numerical modelling, we show that high-frequency acoustic waves ($\omega > \omega_\mathrm{ac}$) undergo constructive and destructive interference within the subsurface cavity between the source and the lower turning point. This naturally generates the observed pattern of pseudomode peaks and troughs in the power spectrum. In contrast to p modes which evanesce above the photosphere, these waves can propagate higher into the solar atmosphere, acting as effective drivers of dynamical processes there \citep[see e.g.][]{2011ApJ...728...84B, 2021ApJ...922..225R}; importantly, this driver exhibits a finely structured spectrum with power concentrated around discrete periods \citep[around 3 minutes and shorter for the Sun, see also][]{2014AstL...40..576Z, 2021A&A...649A.169S, 2025A&A...697A.156B}.
    \item Pseudomode peak frequencies are found to depend on both the photospheric acoustic cut-off frequency $\alpha$ and the source location $r_0$, whereas troughs depend on $r_0$ only. This distinction is potentially testable observationally by comparing the behaviour of individual pseudomode peaks and troughs over the solar cycle. The effective dispersion relation given by Eq.~(\ref{eq:disp_pseudo}) describes the effects of $\alpha$ and $r_0$ on the pseudomode peak frequencies. Thus, the frequency of all pseudomode radial harmonics decreases with $r_0$.
    The effect of $\alpha$ is harmonic-dependent. Variations in $\alpha$ can shift pseudomode peaks either towards higher or lower frequencies, depending on radial harmonic number.
    \item Both $\alpha$ and $r_0$ can, in principle, generate an anti-phase solar-cycle behaviour of pseudomode frequency shifts. While $r_0$-modulation always drives pseudomode frequency in anti-phase with the p-mode frequency variations, $\alpha$-modulation can cause both in-phase and anti-phase responses depending on harmonic order. This offers a potential explanation for why pseudomode frequencies can shift differently across frequency bands for the Sun \citep{rhodes2011temporal} and for different stars \citep{2025MNRAS.537.1268M}. This requires a dedicated follow-up study. Moreover, $r_0$ produces significantly stronger frequency modulation than $\alpha$ (cf. modulation amplitudes in panel (b) and panels (c)-(d) in Fig.~\ref{fig:shifts}), providing a way to distinguish the dominant mechanism observationally. However, a more detailed model is needed for quantitative comparison.
    {Shorter-timescale variations, such as quasi-biennial oscillations, are well known to be also present in p-mode frequency shifts and in other magnetic activity proxies, including radio flux \citep[see e.g.][]{2015MNRAS.451.4360K, 2022MNRAS.515.2415M}. However, investigating the response of pseudomode frequencies to such shorter timescales is beyond the scope of the present study, which is focused on the dominant 11-year modulation, and may represent a potentially interesting direction for future work.}
    \item The model is validated by Bayesian MCMC best-fitting to GONG observations. Indeed, the theoretically derived pseudomode spectrum Eq.~(\ref{eq:fit_model}) reproduces the observed spectrum well {(Fig.~\ref{fig:best-fit})} and yields reasonable values of the model parameters, such as the cavity transit time $\tau_\mathrm{A}\approx43$\,min and depth $a\approx0.5 R_\odot$, source location below the surface $(1-r_0)a\approx 13$\,Mm, and the photospheric cut-off frequency $\alpha\approx4656$\,$\mu$Hz. However, the inferred cavity depth $a$ and the corresponding source height $(1-r_0)a$ appear to be larger than previous estimations, indicating the need for further model development. In addition, we find the power-law index of the superimposed background (parameter $C$ in Eq.~(\ref{eq:fit_model}) and Table~\ref{tab:params}) in the observed spectrum to be about 1.
    {In Fig.~\ref{fig:best-fit}, the GONG data for the mode with $\ell=100$ and $m=0$ was used. The generalisation of this result for multiple modes and the use of data from other existing instruments \citep[e.g. Variability of solar IRradiance and Gravity Oscillations / Sun PhotoMeters, VIRGO/SPM,][]{1995SoPh..162..101F, 1997SoPh..170....1F, 2002SoPh..209..247J} seem to be natural next steps. Also, the exploration of the frequency range within which our model remains capable of making a meaningful pseudomode detection constitutes another interesting follow-up.}
    
\end{itemize}

Our results demonstrate that high-frequency acoustic pseudomodes offer a promising diagnostic probe for the subsurface structure and dynamics of the Sun and other stars. Moreover, the propagating nature and finely structured spectra of pseudomodes make them a natural driver of dynamic processes and activity in the lower solar atmospheric layers \citep[see e.g.][]{2023LRSP...20....1J}.
The combined theoretical and observational approach developed here lays the ground for the seismological use of pseudomodes.
However, while the present cavity model captures the essential physics of pseudomode formation and explains key features of their temporal evolution, further refinements incorporating, for example, realistic $\omega_\mathrm{ac}(r)$ profiles and frequency-dependent turning points, multidimensional effects, and broad-band excitation, are required to unlock the full diagnostic potential of pseudomodes and reconcile theory with observations.

\section*{Acknowledgements}

The work is supported by the STFC consolidated grant ST/X000915/1. DYK also thanks the UKRI Stephen
Hawking Fellowship EP/Z535473/1 and the Latvian Science Council Grant lzp-2024/1-0023.

\section*{Data Availability}
The data underlying this article are available in the article and in the references therein.







%
%
%

\bsp	
\label{lastpage}
\end{document}